# Thermo-optic locking of a semiconductor laser to a microcavity resonance

T. G. McRae[1,2], K. H. Lee[1], M. McGovern[3], D. Gwyther[1] and W. P. Bowen[1]

[1]*Department of Physics, University of Queensland, Brisbane, QLD 4072, Australia*
[2]*MacDiarmid Institute for Advanced Materials and Nanotechnology, University of Otago, Dunedin, New Zealand*
[3]*Jack Dodd Center for Photonics and Ultracold Atoms, University of Otago, Dunedin, New Zealand*

[*]*tmcrae@physics.otago.ac.nz*

**Abstract:** We experimentally demonstrate thermo-optic locking of a semiconductor laser to an integrated toroidal optical microresonator. The lock is maintained for time periods exceeding twelve hours, without requiring any electronic control systems. Fast control is achieved by optical feedback induced by scattering centers within the microresonator, with thermal locking due to optical heating maintaining constructive interference between the cavity and the laser. Furthermore, the optical feedback acts to narrow the laser linewidth, with ultra high quality microtoroid resonances offering the potential for ultralow linewidth on-chip lasers.

16   V. V. Vassiliev, V. L. Velichansky, V. S. Ilchenko, M. L. Gorodetsky, L. Hollberg and A. V. Yarovitsky, "Narrow-line-width diode laser with a high-Q microsphere resonator", Opt. Comm., **158,** 305-312 (1998).

## 1. Introduction

Toroidal microcavities (microtoroids) are of interest for a wide range of fundamental studies and applications, including cavity quantum electrodynamics [1] cavity optomechanics [2] compact optical elements for telecommunications [3] and high-sensitivity chemical/biological sensors [4]. Such interests arise from both the ultra-high quality factors $Q$ ($10^8$) [5] achievable in whispering gallery modes (WGMs), and the lithographic fabrication which leads to natural scalability and straightforward integration into microchip architectures.

Separately, there is a growing need for diode lasers with high spectral purity for high-precision interferometry, high-resolution spectroscopy and metrology. Typically, unstabilized diode lasers have a broad linewidth of around 1-10 MHz due to the low $Q$ of the laser cavity. The use of a wavelength-selective optical feedback is a common technique for linewidth narrowing and wavelength tuning [6]; with the feedback achieved by a number of methods including resonant reflection from a confocal cavity [7], or back-scattering from a fiber-optic cavity [8], or microsphere resonator [9]. A real issue in optical feedback based laser locking techniques is maintaining constructive interference between the cavity field and the feedback field. This is usually achieved with electronic feedback control which adds significant complexity.

In this Letter we demonstrate a new application of microtoroids as a wavelength-selective optical feedback element for semiconductor lasers by taking advantage of the back-scattered light from the microtoroid. This backscattered light is the result of surface imperfections and the long photon lifetimes in microtoroids which cause strong modal coupling between counter propagating modes [10]. Here we show that feedback both reduces the semiconductor linewidth and improves the frequency stability over both very short and very long time intervals. Furthermore we utilize the thermal bistability present in silica microtoroids [11] to achieve thermo-optic locking between the laser and cavity for periods surpassing 12 hours, eliminating the requirement of electronic servo control in typical optical feed-back based laser stabilization techniques.

## 2. Experiment Description

Figure 1 illustrates our experimental setup. Briefly, a 5mW laser diode is coupled into a tapered optical fiber and used to excite a microtoroid cavity; with a reference laser used to both characterize the locking and locate suitable microtoroid resonances. The laser diode (ML925B45F) had no antireflection coating on the output facet, and its frequency was tuned with current and temperature control. The power and polarization of the excitation field was controlled with an attenuator, and quarter ($\lambda/4$) and half ($\lambda/2$) wave plates, respectively. The semiconductor laser power was coupled into the tapered fiber through a fiber coupler at point "A" in Fig. 1. Switching the input to the fiber coupler allowed us to switch between the feedback experiment, and the use of the reference laser to characterize cavity resonances. To avoid parasitic backscattering from the taper we used a high efficiency (95%) adiabatic tapered optical fiber to couple the evanescent fields of the taper and the microtoroid. The microtoroid was positioned on a piezoelectric actuator which provided precise control of the coupling distance, and hence the coupling strength. The feedback field was split with a 50/50 fiber coupler to monitor the feedback into the semiconductor laser. The transmitted (forward-propagating) and feedback (backward-propagating) fields were then detected with New Focus 125-MHz photoreceivers. The microtoroids used in this study were manufactured with the same fabrication process as described in [5], with each chip consisting of forty microtoroids. The reference laser, a single mode semiconductor laser, (New Focus Velocity®

6328) had its own isolation and polarization control and was able to scan the large free spectral range (ca 11nm) of the cavity without mode hopping.

First we characterize the WGM of the microtoroid used to provide the feedback signal. Figure 2(a) shows the transmitted and feedback signals from the cavity for a typical resonance at 1560.4nm in the critical coupling regime, while Fig. 2(b) shows the same WGM in the under-coupled regime, where the tapered fiber has been pulled far from the cavity. In the under-coupled regime we see that the resonance begins to split into a doublet due to the coupling between the two counter propagating modes inside the cavity [10,12] This mode splitting is a prerequisite for optical locking, indicating that the backscattering rate is larger than the loss rate of the system, and hence that significant backscattered power can be achieved. The model of [13,14] was used, with good agreement, to fit these observations and gave a $Q$ of $3.1 \times 10^6$.

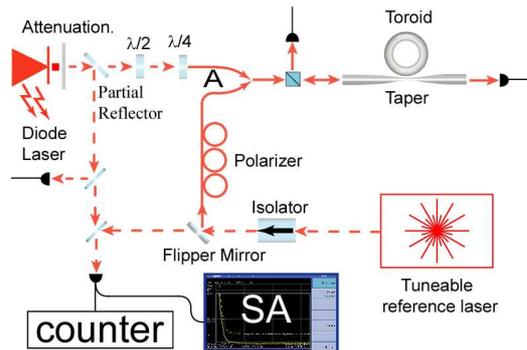

Fig. 1. (Color online.) Schematic of the experimental setup. Solid and dashed red lines respectively show optical paths in fiber and in free space. "A" indicates the point where switching between the reference laser and the semiconductor laser is performed.

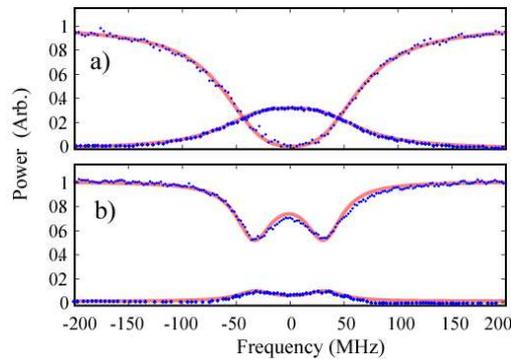

Fig. 2. (Color online) Spectral response of the forward and backward propagating fields in the microtoroid in the critically coupled and undercoupled regimes. (a) Critically coupled. Top curve: forwards propagating field. Bottom curve: backwards propagating field. Solid lines: theoretical fit. (b) Undercoupled linewidth.

## 3. Thermo-optic Locking

The well known sensitivity of semiconductor lasers to optical feedback is a result of the very flat gain curve as a function of wavelength, the short cavity with low finesse, and the fact that when light is returned to the cavity the laser acts as a photodetector generating more carriers across the junction and affecting net laser gain [6]. The laser sees optical feedback when its frequency matches a resonance of the microtoroid. As a consequence the coupled system (laser and microtoroid) experiences highest gain near cavity resonance. Hence laser emission grows fastest there, quenching emission at other frequencies and automatically locking the laser frequency onto resonance. The optical feedback provides fast frequency corrections to the laser that dramatically narrow the linewidth, as shown in Fig. 5. However optical locking alone does not solve the problem of long term stability. The path length between the laser and the feedback element fluctuates over time scales longer than milliseconds due to environmental factors such as mechanical vibrations, thermal effects and air turbulence. This alters the phase of the feedback into the laser diode. If the interference between the feedback and the field in the laser diode becomes destructive, the laser will lose lock.

To achieve long term stability we use the microtoroids own thermal response to preserve the constructive feedback required to keep the laser on cavity resonance. This thermal response has been used previously to lock a microtoroid onto a laser isolated from optical feedback [15]. In this case, optical heating due to light within the resonator causes thermal expansion and a corresponding shift in the microtoroid resonance frequency. In silica microtoroids the dominant effect is due to the temperature dependent refractive index of the material which results in a negative frequency shift with increasing temperature. Locking is achieved in a region where the resonator frequency is detuned below that of the laser. In this region, heating due to optical excitation of a cold resonator pushes the resonator frequency further from the laser frequency, and hence reduces the intracavity power and decreases the heating effect. At sufficient detuning, the rate of heat loss to the environment balances the optical heating and a stable equilibrium is established

In our case, the situation is more complex, with competition between the thermal and optical feedback based locking processes. The optical feedback locking optimizes the laser frequency with respect to the microtoroid resonance to maximize the constructive interference between the feedback and intracavity laser fields; whilst the thermal locking optimizes the microtoroid resonance frequency with respect to the laser frequency to balance optical heating and thermal heat loss. Since the optical locking occurs over timescales fast compared to the microtoriod thermal decay time, we can model this process assuming that the optical feedback locking dominates and the laser frequency is always such that the constructive interference between feedback and intracavity laser fields is maximized. To understand the combined locking system, consider the case where the optical path length between toroid and laser is optimized to maximize the constructive interference at the cold microtoroid resonance frequency. On turn-on, the laser will immediately oscillate at this frequency, with the feedback power maximized. However, optical heating will then drag the microtoroid resonance frequency lower, both decreasing the feedback power and degrading the constructive interference at the microtoroid resonance frequency. The laser frequency is consequently also pulled lower, but no longer aligns exactly with the microtoroid resonance since improved constructive interference can now be achieved off resonance. The result over time is a net negative shift in the laser frequency, and an exponential decay of both the microtoroid intracavity power and the feedback power; until, as with thermal locking alone, the optical heating and thermal heat loss exactly balance and a stable equilibrium is established. Fig. 3(a) shows experimental measurements of the exponential decay of the feedback power into equilibrium, giving a thermal time constant of about ~40µs. Fig. 3(b-d) present models of the thermal pulling process which are well matched to the experimental observations.

**4. Results**

The power of the forwards and backwards propagating fields as the tapered optical fiber is brought towards the toroid is shown in Fig. 4, with the free running laser frequency set to near the toroid resonance. One observes discrete power steps, as the semiconductor laser locks to different toroid modes, with the backwards propagating light jumping from zero power initially to a maximum of 18µW at optimal coupling. Once optimal coupling was achieved, the taper position was held fixed, with the laser maintaining thermo-optic lock for periods upwards of 12 hours and both forwards and backwards propagating powers maintained stably, with fluctuations due only to only to variation of the tapered optical fiber position. Figure 4(c) shows the power reflected from the partial reflector with 10% reflectivity shown in Fig 1. The linewidth of this beam is narrowed in the same way as those interacting with the toroid, however its intensity is insensitive to taper position fluctuations and therefore provides a highly intensity stable output. This output was used to characterize the linewidth of the stabilized laser through a beat frequency measurement with the reference laser. The beat frequency was measured on a 1GHz bandwidth detector and monitored on a spectrum analyzer (Agilent N9320A). Its Allan variance was determined with a frequency counter (Stanford Research Systems SR620). Figure 5(a) shows the beat signal for the free running laser, while Fig. 5(b) shows the beat signal of the laser when in thermo-optic lock. The locked semiconductor laser runs single mode as indicated by the presence of a single peak in the frequency spectrum. The most narrow linewidth observed from the free running laser was 1.4 MHz full width at half maximum (FWHM), whereas the locked laser has an observed linewidth of 300 kHz (FWHM) limited by the linewidth of our reference laser. By comparison with other WGM based optical locking experiments, it is reasonable to expect that the actual locked laser linewidth to be significantly lower [9,16].

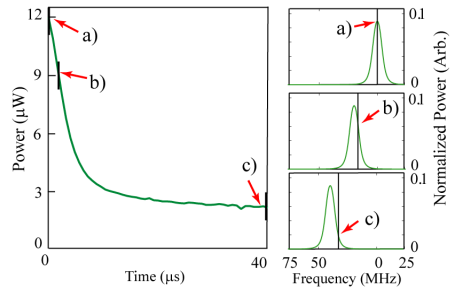

Fig. 3. Measurement and model of thermo-optic locking process for a toroid with a $Q$ of $1\times10^7$ and a scattering rate of 8MHz. (Main figure) Measurement showing the back reflected power immediately after obtaining lock. (a-c) Model of the thermally shifted microtoroid spectral response with the laser locking frequency indicated by the vertical line, corresponding to the three labeled points in the experimental data.

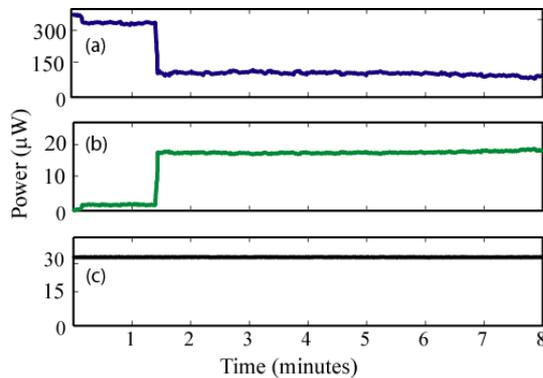

Fig 4.(Color online.)Typical trace of the laser acquiring lock. (a) Forwards propagating power. (b) Backwards propagating power. Fig. (c) Power tapped off from partial reflector.

The beat note Allan Variance measurements allowed a frequency stability comparison to be made between the locked and free running lasers. Figure 5(c) shows the Allan variance for the free running and locked semiconductor lasers over a 12 hour period. It is clear that over time scales less than about 5 seconds the stability of the locked laser, with the narrow linewidth, surpasses that of the free running laser. Over relatively long times, greater than thirty minutes, the locked laser also outperforms the monotonic frequency drift of the unlocked laser. Between these time scales the free running laser outperformed the thermo-optically locked semiconductor laser. The degradation in the thermo-optically locked lasers performance is due to fluctuation in the path length between laser diode and the microtoroid. As the path length fluctuates, the lock adjusts the laser frequency to maintain constructive feedback. One can envisage overcoming this issue with a system of the type proposed by [16] where the microtoroid-laser diode distance is on the order of several millimeters. Indeed the whole laser waveguide and toroid system could be fabricated on the same thermally stabilized chip. This should dramatically reduce the Allan variance of the locked laser across the whole frequency spectrum.

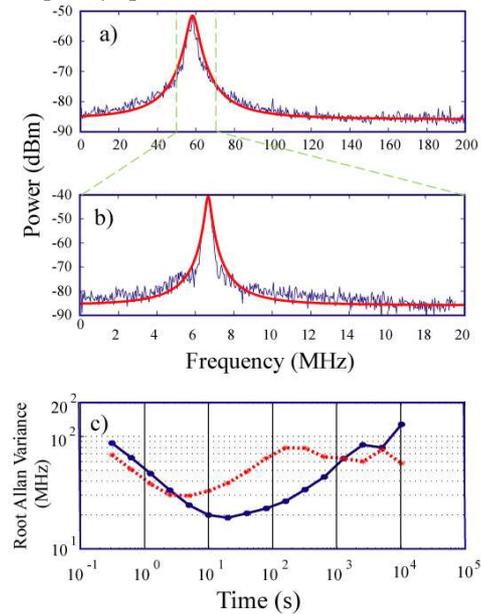

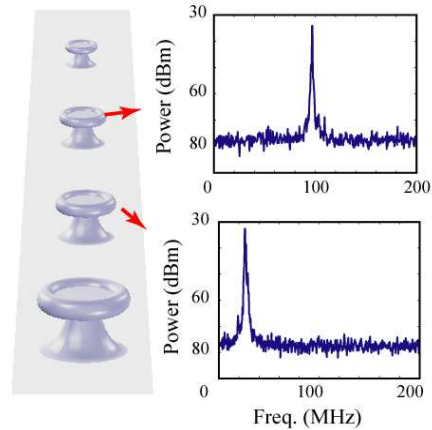

Fig. 6. (Color online) Schematic of single chip with multiple integrated microtoroid resonators each enabling low linewidth frequency lasers at different wavelengths.

Fig 5. (Color online.) Comparison of beat measurements for thermo-optically locked and free running lasers. (a)Beat signal for locked laser. (b)Most narrow beat for unlocked laser. Solid lines: Lorentzian fit. (c)Allan variance of beat note frequency. Locked laser red dashed line. Unlocked laser solid blue line.

**Conclusion**

In conclusion, we have demonstrated a new type of wavelength-selective thermo-optic feedback element based on microtoroid resonators. We demonstrate stable locking over twelve hours with a reduction in linewidth from 1.4 MHz to 300 kHz, limited by measurement resolution. The locking is achieved without any electronic locking techniques, has immediate applications in areas which require intra cavity enhancement of optical fields, such as non linear optics based experiments and optical resonator based biological or chemical sensors [4]. Such a device could enable low cost very low linewidth lasers and, in future, a high stability, low cost frequency reference. The "on-chip" architecture provides scope for integration and with an array of microtoroids providing a range of output wavelengths as shown in Fig. 6.

**Acknowledgements**

We thank Joachim Knittel for useful discussions. This research was supported by the Australian Research Council Discovery Project DP0987146 and the New Zealand Foundation for Research Science and Technology under the contracts NERF-UOOX0703: Quantum Technologies and NERF-C08X0702: Theory and Applications for Communications Technologies.